\documentclass[aps,prb,twocolumn,superscriptaddress,10pt]{revtex4-2}
\usepackage[utf8]{inputenc}
\usepackage{libertine}
\usepackage[libertine]{newtxmath}
\usepackage{mathtools,graphicx,microtype,bm}
\usepackage[cal=boondoxo,frak=boondox]{mathalfa}
\usepackage[colorlinks=true,allcolors=blue]{hyperref} 

\begin{document}
\title{Friedel oscillations in two-dimensional materials with inverted bands and Mexican-hat dispersion}

\author{Vladimir A.\ Sablikov}
\email[E-mail:]{sablikov@gmail.com} 
\affiliation{Kotelnikov Institute of Radio Engineering and Electronics, Fryazino Branch, Russian Academy of Sciences, Fryazino, Moscow District, 141190, Russia}

\begin{abstract}
We study Friedel oscillations (FOs) in two-dimensional topological materials with Mexican hat band dispersion, which attract great interest due to the bunch of its inherent non-trivial features, including the Van Hove singularity, doubly connected Fermi surface, non-trivial quantum-geometric properties, and the presence of states with negative effective mass. These factors are found to lead to a three-mode structure of the FOs. One of the modes, arising from electron transitions between the Fermi contours, has an unexpectedly large amplitude. The evolution of the amplitudes of all modes with Fermi energy is largely determined by the interplay of three main factors: intra-contour and inter-contour electron transitions, the quantum metric of the basis states, and the electron-electron interaction. We traced the role of each factor in the formation of the FO pattern and identified the corresponding features of the FO evolution.
\end{abstract}

\maketitle

\section{Introduction}\label{S_intro}
Friedel oscillations (FOs) are a fundamental feature of the density-density response specific to a fermionic system. The FOs are caused by the interference of quantum states scattered by a local potential perturbation created by an external charge, defect, interface, etc.~\cite{doi:10.1080/14786440208561086}. Being an essential component of the density-density response function, the FOs reflect many properties of the system of fermionic quasiparticles~\cite{PhysRevB.72.045127,PhysRevB.93.205117}, including their quantum geometry, valley and orbital structure, which are currently of great interest. This allows one to consider FOs as a promising tool for studying non-trivial properties of electronic states using scanning tunneling microscopy (STM). Thanks to advances in STM, the FO concept has evolved in recent decades as a methodological basis for the thriving field of quasiparticle interference (QPI) research. The QPI method has led to significant advances in the study of electronic states in a wide range of materials, from superconductors~\cite{doi:10.1126/science.1072640} to modern quantum materials and correlated states~\cite{Chen_2017,https://doi.org/10.1002/adma.201707628}.

This perspective stimulated numerous theoretical studies of specific features and properties of the FOs for electron systems in graphene-based materials~\cite{PhysRevLett.97.226801,PhysRevLett.101.156802}, topological insulators~\cite{PhysRevB.89.195417,PhysRevB.107.035305}, transition-metal dichalcogenides~\cite{Chen_2017,PhysRevB.97.205410}, Weyl semimetals~\cite{PhysRevB.86.195102,PhysRevB.109.035145}, two-dimensional (2D) electron systems with spin-orbit interaction~\cite{PhysRevB.74.045307,PhysRevB.81.205314}. It was found that in electron systems with parabolic dispersion, the Coulomb interaction between electrons affects the shape of the FOs at relatively short distances, but asymptotically, the FOs are described by doubling the Fermi wave vector~\cite{PhysRevB.72.045127}. Anisotropy of the effective mass leads to a corresponding anisotropy of the FOs~\cite{PhysRevB.103.165303}. Spin-orbit interaction radically changes the wave vector of the FOs~\cite{PhysRevB.74.045307}, and in the presence of two types of spin-orbit interactions, beatings of the FOs arise~\cite{PhysRevB.81.205314}. Multimode FOs and the beatings of electron density appear in the presence of valleys in the band spectrum~\cite{lin2024multimode}.

Recently, there has been increasing interest in the question of what role quantum geometric properties of eigenstates play in the density-density response and related phenomena such as plasma excitations~\cite{PhysRevB.106.155422,PhysRevB.110.045403} and FOs~\cite{PhysRevLett.125.116804,PhysRevB.104.035402,https://doi.org/10.1002/pssr.202300378}. The significance of this problem is largely determined by the discovery of striking manifestations of quantum geometry in the structure of the FOs, which opens up promising prospects for the experimental study of geometric properties. One way to implement this idea is to measure the Berry phase by observing the wavefront dislocations in the FOs created by a charge embedded in the host crystal~\cite{dutreix2019measuring,PhysRevLett.125.116804,PhysRevB.104.035402,liu2024visualizing}. Another possibility is to study the characteristic features of the FOs that arise due to the quantum geometry of band states. In this case, the question first of all arises about the features of the spatial shape of the FOs and their behavior when changing the parameters of the system, in particular, the Fermi energy. The research presented here is aimed precisely in this direction.

The key moment determining the geometric properties and topology of quantum states is the multi-orbital structure of the Bloch states~\cite{PhysRevLett.131.240001}, i.e.\ the presence in their composition of several orbitals with different properties. The multi-orbital structure of the wave functions manifests itself in significant and even dramatic changes of both charge screening and plasma excitations, as studies have shown for some topological materials~\cite{PhysRevLett.119.266804}, graphene~\cite{PhysRevB.106.155422}, and $\alpha-\mathcal{T}_3$ lattice structure~\cite{PhysRevB.106.155422,iurov2021tailoring}.

In this paper, we study the FOs for 2D materials with Mexican hat-shaped band dispersion formed by the inversion of electron and hole bands. This system is interesting for the following reasons. First, the basis quantum states are topological with a sufficiently large Berry curvature. Second, in the energy range between the bottom and the top of the Mexican hat dispersion (MHD), there are two Fermi contours. Third, the MHD has two important nontrivial features. The main one, which usually attracts attention, is the Van Hove singularity of the density of states at the MHD bottom. Because of this, the role of electron–electron (e-e) interaction increases significantly and, as a consequence, the conditions for the formation of the ferromagnetic phase~\cite{PhysRevB.75.115425,PhysRevLett.114.236602} and superconducting pairing~\cite{PhysRevLett.56.2732,PhysRevLett.98.167002} are facilitated. Another feature is related to the effective mass of quasiparticles. On the low-wave vector branch of the MHD, the effective mass changes sign with increasing energy from positive near the MHD bottom to negative near the top. This obviously affects the distribution of electron density around the external charge and, consequently, the screened potential, since quasiparticles with negative mass are attracted to the negatively charged center. In particular, due to this feature, quasi-bound states with energy above the MHD top are formed~\cite{SABLIKOV2023115492,SABLIKOV2023129006}.

Thus, one can expect dramatic changes in the spatial distribution of electron density around an external charge. Since MHD occurs in many 2D electron systems, this problem is of rather general interest.

We found that three Kohn anomalies are formed due to the presence of two Fermi contours. In addition to the two anomalies caused by intra-contour transitions, there is another anomaly caused by inter-contour transitions. At low Fermi energies, the inter-contour anomaly significantly exceeds the other two. The quantum metric significantly affects the amplitude of the singularities. In particular, the inter-contour singularity disappears altogether at high Fermi energies. As a result, a three-mode FO structure is formed, the evolution of which with the Fermi energy is determined by the interaction of three main factors: intra-contour and inter-contour transitions, the quantum metric, and the e-e interaction, which plays an important role due to the singularity of the density of states.

The calculations are performed in the random phase approximation (RPA) for 2D electron systems described by the BHZ model. In Sec.~\ref{S_Lindhard}, we study the Lindhard polarization function. The screened potential of a point charge and the electron density oscillations around it are studied in Sec.~\ref{S_Friedel}. In Sec.~\ref{S_discuss} we discuss the main results and draw conclusions.

\section{Lindhard polarization function}\label{S_Lindhard}

The electron density response to an external potential is calculated in the RPA, which is commonly used for this purpose. In this approximation, the key role is played by the Lindhard polarization function, which describes the density-density response of non-interacting electrons~\cite{giuliani2008quantum}. As a function of the wave vector $\bm{q}$ and the frequency $\omega$, the Lindhard polarization function $\Pi^{(0)}(\bm{q},\omega)$ reads~\cite{giuliani2008quantum,PhysRevLett.119.266804}
\begin{equation}\label{eq.Lidhard_fun}
    \Pi^{(0)}\!(\bm{q},\omega)=\sum_{\lambda,\lambda'}\int\! \frac{d^2\bm{k}}{(2\pi)^2} \frac{n(E_{\lambda,\bm{k}})-n(E_{\lambda',\bm{k}+\bm{q}})}{\hbar \omega +E_{\lambda,\bm{k}}-E_{\lambda',\bm{k}+\bm{q}}+i\eta} \mathcal{F}_{\lambda,\lambda'}(\bm{k},\bm{k}+\bm{q})\,,
\end{equation}
where $E_{\lambda,\bm{k}}$ is the energy of the band state
\begin{equation}
    |\lambda, \bm{k}\rangle =\frac{1}{L}u_{\lambda}(\bm{k})e^{i\bm{k r}}\,,
\end{equation}
with wave vector $\bm{k}$. Here $\lambda$ is the band index and/or other discrete quantum number of the state (including spin $s$ if it is conserved) and $u_{\lambda}(\bm{k})$ is a spinor whose rank is determined by the spin and atomic orbitals that form the eigenstates, $n(E_{\lambda,\bm{k}})$ is the occupation number, $L$ is a normalization length. 

The form factor $\mathcal{F}_{\lambda,\lambda'}(\bm{k},\bm{k}+\bm{q})$ describes the overlap between the cell periodic parts of the Bloch eigenstates with different quantum numbers:
\begin{equation}\label{eq.overlap_func}
    \mathcal{F}_{\lambda,\lambda'}(\bm{k},\bm{k}+\bm{q})=|u^+_{\lambda}(\bm{k})\, u_{\lambda'}(\bm{k}+\bm{q})|^2\,.
\end{equation}
This function is directly related to the quantum metric of eigenstates~\cite{resta2011insulating}, $D_{\lambda,\lambda'}(\bm{k},\bm{k'})^2=1-|u^+_{\lambda}(\bm{k}) u_{\lambda'}(\bm{k'})|^2$. Effects due to the quantum metric are usually studied in the long-wave approximation, for example, for plasmons, when the overlap function is expanded in $q$~\cite{PhysRevB.110.045403}. An important feature of our situation is that the dependence of the quantum metric on $k$ and $q$ is of great importance when $q$ is close to the poles of the integrand in the Lindhard function, Eq.~(\ref{eq.Lidhard_fun}), when the overlap function can significantly change the value of the integral and even affect the presence of the Kohn anomaly.

The MHD in 2D topological insulators can be described within the well-known BHZ model~\cite{BHZ}. In this model, the eigenstates are formed as a result of $sp^3$ hybridization of the electron and hole bands. If the inversion symmetry is not broken, the spin component perpendicular to the layer is a good quantum number, and the total Hamiltonian splits into two Hamiltonians for each spin orientation. In this case, it is convenient to replace the index $\lambda$ with two indices: the band index $\lambda=\pm 1$ and the spin index $s=\uparrow, \downarrow$. The spin-up Hamiltonian is~\cite{BHZ}
\begin{equation}\label{eq.Hamiltonian}
    H_{\uparrow}= 
    \begin{pmatrix}
        -M+B \hat{k}^2 & A (\hat{k}_x+i\hat{k}_y)\\
        A (\hat{k}_x-i\hat{k}_y) & M-B \hat{k}^2\\
    \end{pmatrix},
\end{equation}
where for simplicity we assume that the electron and hole bands are symmetric. $M$, $B$ and $A$ are standard parameters of the model. Their numerical values are known for various materials.

In what follows we turn to dimensionless quantities. The values of the energy dimension are normalized to $|M|$, the distance is normalized to $\sqrt{|B/M|}$, the wave vector $k$ is normalized to $\sqrt{|M/B|}$. An important parameter of the model is $a=A/\sqrt{B M}$. It describes the hybridization of the electron and hole bands. The MHD is realized when $|a|<\sqrt{2}$. The dispersion relation is $\varepsilon(\lambda, k)=\lambda\varepsilon_k$, with
\begin{equation}    
\varepsilon_k=\sqrt{(1-k^2)^2+a^2k^2}\,.
\end{equation}

The spinor $u_{s, \lambda}(\bm{k})$ for spin-up states has the form
\begin{equation}\label{eq.u-spinor}
    u_{\uparrow, \lambda}( \bm{k}) = \frac{1}{\sqrt{1+\beta_{\lambda, \bm{k}}^2}}
    \begin{pmatrix}
        1 \\ \beta_{\lambda, \bm{k}} e^{-i\phi}  
    \end{pmatrix}\,,
\end{equation}
where 
\begin{equation}
    \beta_{\lambda, \bm{k}}=\frac{a k}{\lambda\varepsilon_k-1+k^2}\,
\end{equation}
and $\phi$ is the polar angle of the vector $\bm{k}$.

The Berry curvature of the states $|s,\lambda,\bm{k}\rangle$ is equal to
\begin{equation}
    \bm{\Omega}_{s,\lambda}(\bm{k})=-s\lambda\frac{a^2(1+k^2)}{2\varepsilon_k^3}\bf{e}_z\,,
\end{equation}
where spin index $s=\pm 1$.

\begin{figure}
    \centerline{\includegraphics[width=1.\linewidth]{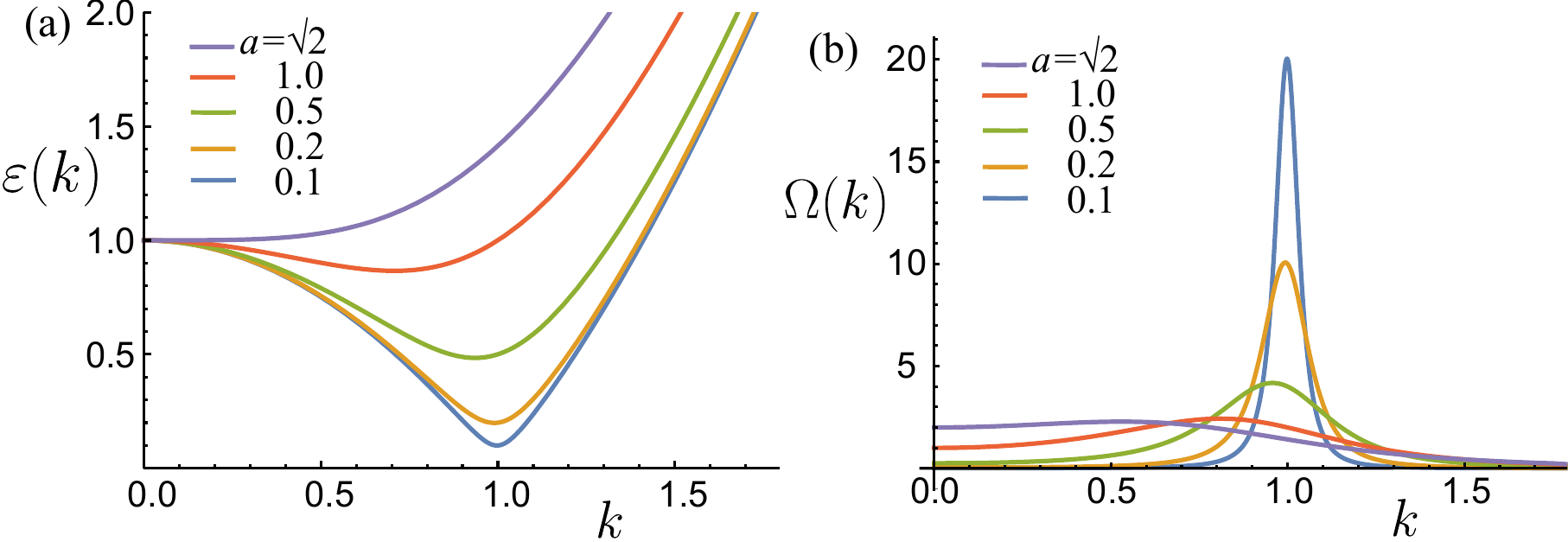}}
    \caption{Evolution of (a) the energy dispersion and (b) the Berry curvature with changing the hybridization parameter $a$ in the range $0<|a|<\sqrt{2}$, where MHD exists.}
    \label{fig1}
\end{figure}

The evolution of the energy dispersion $\varepsilon_k$ with the parameter $a$ and the corresponding change in the Berry curvature $\Omega_{\uparrow,+}(k)$ are shown in Fig.~\ref{fig1}. The MHD bottom is at $\varepsilon_c=|a|\sqrt{1-a^2/4}$, the top is at $\varepsilon_t=1$. For $a\ll 1$, the Berry curvature is large and has a sharp peak as a function of $k$. As $a$ increases, the peak decreases and expands, but the integral of $\Omega_{s,\lambda}(\bm{k})$ over the band, of course, remains constant.

Quantum metric effects in the density-density response are determined by the overlap function~(\ref{eq.overlap_func}). Straightforward calculation  within the BHZ model shows that $\mathcal{F}_{\lambda,\lambda'}(\bm{k},\bm{k}+\bm{q})$ has the form
\begin{equation}\label{eq.F_1,2}
    \mathcal{F}_{\lambda,\lambda'}(\bm{k},\bm{k}+\bm{q})=\frac{1}{2}+\lambda \lambda' \Delta\mathcal{F}(\bm{k}\bm{q})\,,
\end{equation}
where
\begin{multline}
     \Delta\mathcal{F}(\bm{k}, \bm{q})=[(1-k^2)(1-k^2-q^2)+a^2k^2\\+(a^2-2+2k^2)kq \cos(\phi-\theta)]/[2\varepsilon_k \varepsilon_{|\bm{k}+\bm{q}|}], 
\end{multline}
with $\theta$ being the angular coordinate of the vector $\bm{q}$.

\begin{figure}
    \centerline{\includegraphics[width=1.\linewidth]{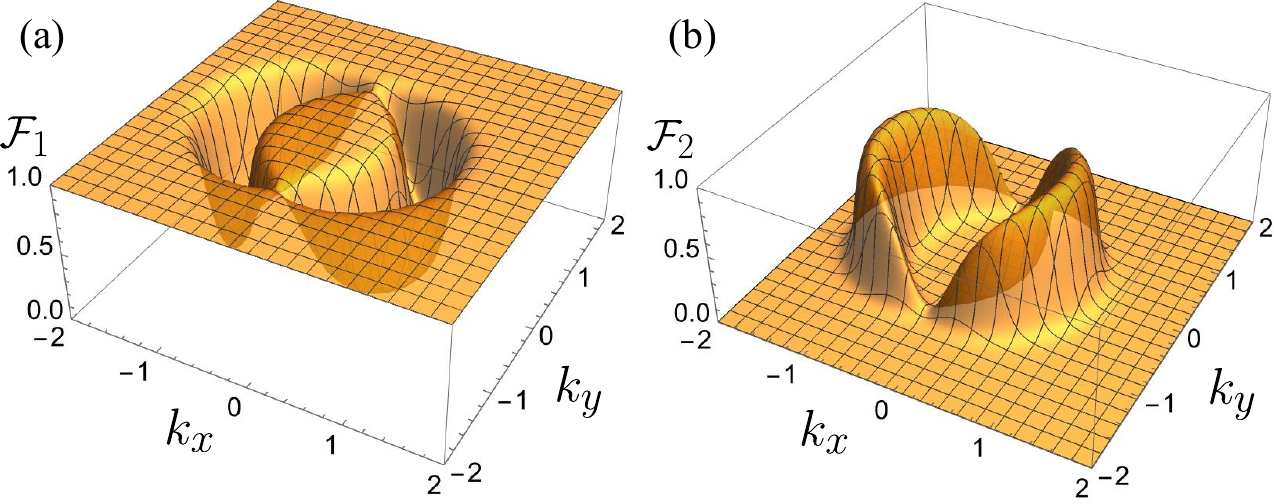}}
    \caption{landscape of the overlap functions $\mathcal{F}_{1, 2}$ for (a) intraband and (b) interband transitions in $k$-space. The scattering wave vector and the hybridization parameter are fixed, $q=0.5$ and $a=0.2$.}
    \label{fig2}
\end{figure}

Thus, the matrix $\mathcal{F}_{\lambda,\lambda'}$ contains on two independent elements, a diagonal $\mathcal{F}_1=\frac{1}{2}+\Delta\mathcal{F}$ and an off-diagonal $\mathcal{F}_2=\frac{1}{2}-\Delta\mathcal{F}$, corresponding to intraband ($\lambda'=\lambda$) and interband ($\lambda'=-\lambda$) transitions. As a function of $\bm{k}$, each of them varies in the interval from zero to one. An example demonstrating the behavior of the functions $\mathcal{F}_{1, 2}$ in $\bm k$-space is shown in Fig.~\ref{fig2}, where their 3D plots are presented for given values of $q$ and $a$. As $q$ increases, the width and depth of the valleys of the $\mathcal{F}_1$ landscape increase, and the hills of the $\mathcal{F}_2$ landscape change accordingly.

The density-density response for each of the spin components is split according to Eq.~(\ref{eq.F_1,2}) into two parts corresponding to the intraband and interband transitions:
\begin{equation}\label{eq.Pi(0)}
    \Pi^{(0)}(\bm{q}, \omega)=\Pi^{(0)}_{\rm{intra}}(\bm{q}, \omega)+\Pi^{(0)}_{\rm{inter}}(\bm{q}, \omega)\,,
\end{equation}
where 
\begin{equation}\label{eq.Pi_1}
    \begin{split}
     \Pi^{(0)}_{\rm{intra}}(\bm{q}, \omega)=\int\frac{d^2k}{(2\pi)^2} \mathcal{F}_1(\bm{k}, \bm{k}+\bm{q})\left[n(\varepsilon_k)+1-n(-\varepsilon_k)\right] \\ \times \frac{\varepsilon_k-\varepsilon_{|\bm{k}+\bm{q}|}}{(\varepsilon_k-\varepsilon_{|\bm{k}+\bm{q}|})^2-(\omega +i\eta)^2}\,,   
    \end{split}
\end{equation}
and
\begin{equation}\label{eq.Pi_2}
    \begin{split}
     \Pi^{(0)}_{\rm{inter}}(\bm{q}, \omega)=\int\frac{d^2k}{(2\pi)^2}\mathcal{F}_2(\bm{k}, \bm{k}+\bm{q})\left[n(\varepsilon_k)-n(-\varepsilon_k)\right] \\ \times \frac{\varepsilon_k+\varepsilon_{|\bm{k}+\bm{q}|}}{(\varepsilon_k+\varepsilon_{|\bm{k}+\bm{q}|})^2-(\omega +i\eta)^2}\,.   
    \end{split}
\end{equation}

Now we turn to the Lindhard function in the static case $\Pi^{(0)}(q,0)$ at zero temperature. The analysis is greatly simplified since the function $\Pi^{(0)}(\bm{q}, \omega)$ does not depend on the angle $\theta$. This is easy to see from Eqs.~(\ref{eq.Pi_1}),~(\ref{eq.Pi_2}) after integration over the angular variable $\phi$. 

Kohn anomalies are known to arise at such values of $q$ when different Fermi contours have a point of contact. At these points, the integrand in Eq.~(\ref{eq.Pi_1}) for intraband transitions has an integrable divergency. Therefore, the value of the integral is largely determined by the behavior of the overlap function in a vicinity of these points. It is for this reason that the overlap function determines the amplitude of the Kohn anomaly and even its presence.

In the case of the MHD, the Fermi contours are two circles of the radius $k_{F1}$ (inner) and $k_{F2}$ (outer), the space between which is filled with electrons. Accordingly, one can expect that the Kohn anomaly will appear at three values of $q$: $q_0=k_{F2}-k_{F1}$, $q_1=2k_{F1}$, and $q_2=2k_{F2}$. 

Direct calculations using Eq.~(\ref{eq.Pi(0)}) show that in reality the situation with anomalies is more complicated. In general, there are three anomalies, one of which, at $q_0$, is due to inter-contour transitions, and the other two, at $q_1$ and $q_2$, are due to intra-contour transitions. However, some anomalies can disappear due to the quantum metric of the basis states, which is represented by the overlap function. This happens under certain conditions, which are determined by the Fermi energy and the e-e interaction parameter. In addition, the amplitudes of all anomalies change significantly due to the overlap function. 

\begin{figure}
    \centerline{\includegraphics[width=0.9\linewidth]{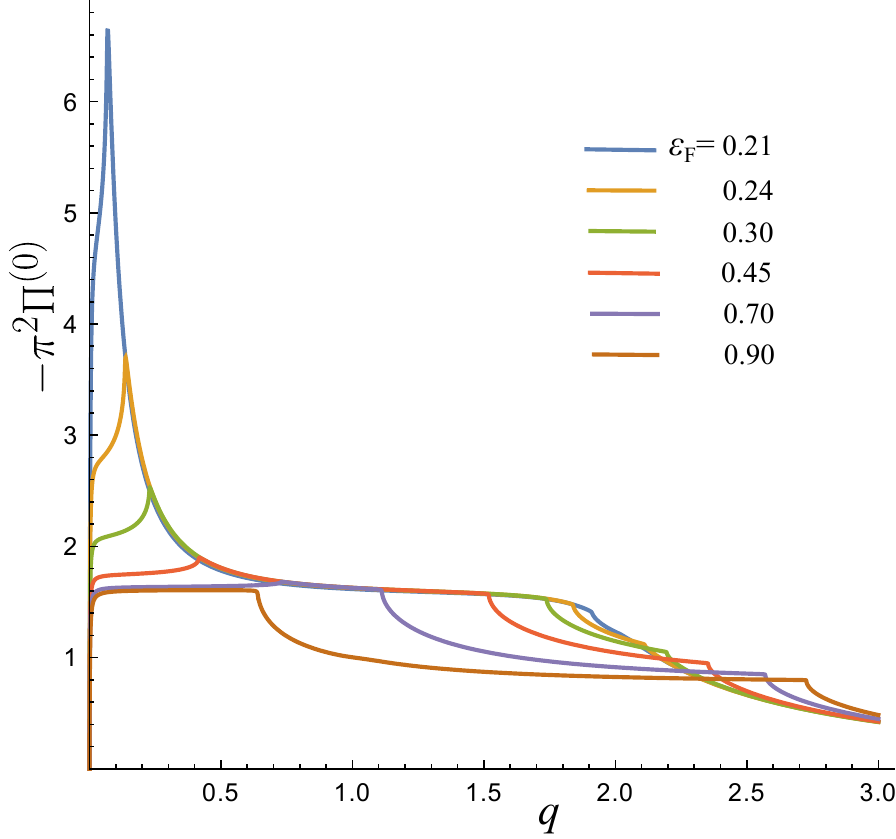}}
    \caption{Lindhard function for a set of Fermi energies from the bottom to the top of the MHD.\@ Hybridization parameter a=0.2.}
    \label{fig3}
\end{figure}

\begin{figure}
    \centerline{\includegraphics[width=0.9\linewidth]{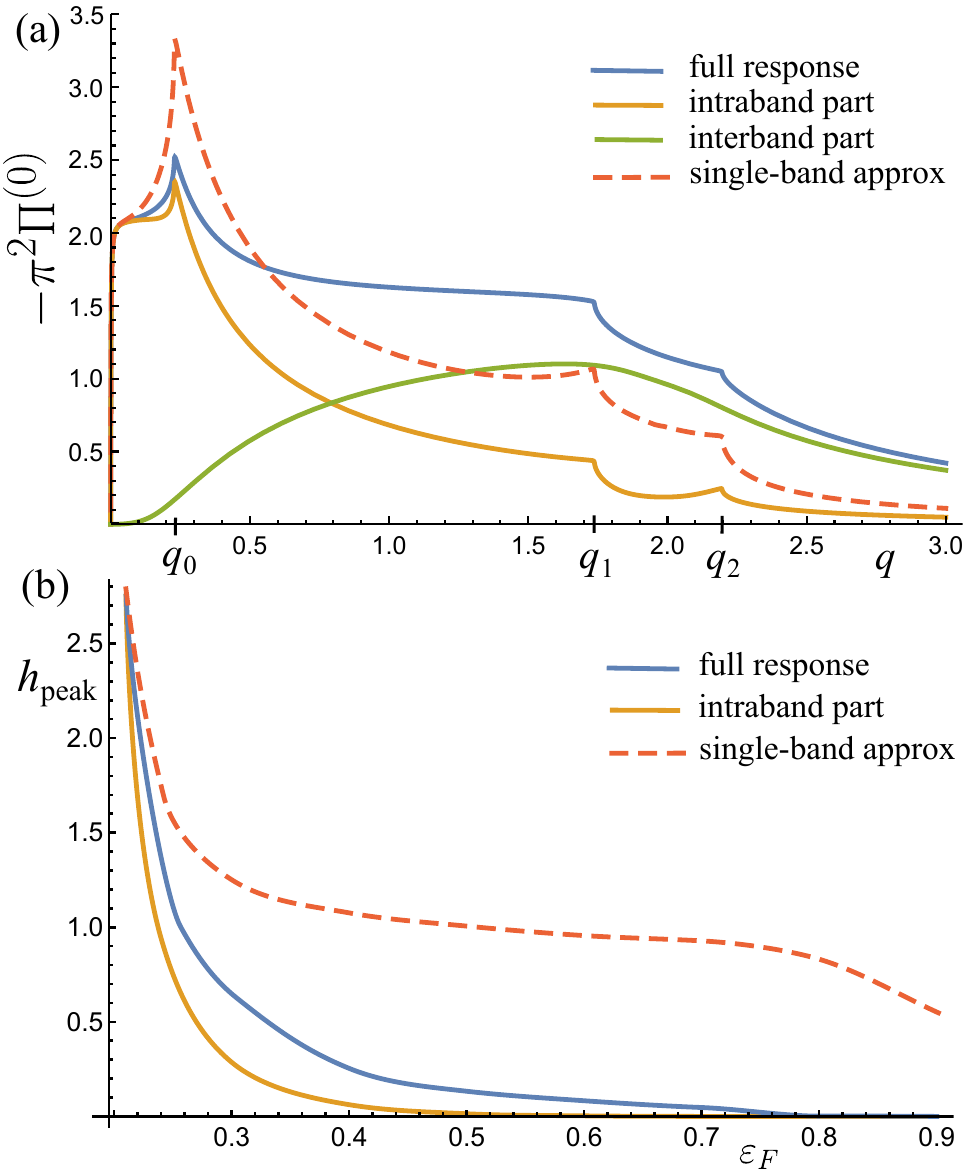}}
    \caption{(a) The full Lindhard function (blue line), its parts due to intraband and interband transitions (yellow and green lines), and ``single-band'' Lindhard function (dashed line) for $\varepsilon_{F}<\varepsilon_{F}^*$. Calculations are carried out for $a=0.2$ and $\varepsilon_{F}=0.3$, when $\varepsilon_{F}^*\approx 0.8$. (b) Peak height of the full Lindhard function, its intraband part, and ``single-band'' Lindhard function depending on $\varepsilon_F$ for the $q_0$-singularity.}
    \label{fig4}
\end{figure}

\begin{figure}
    \centerline{\includegraphics[width=0.9\linewidth]{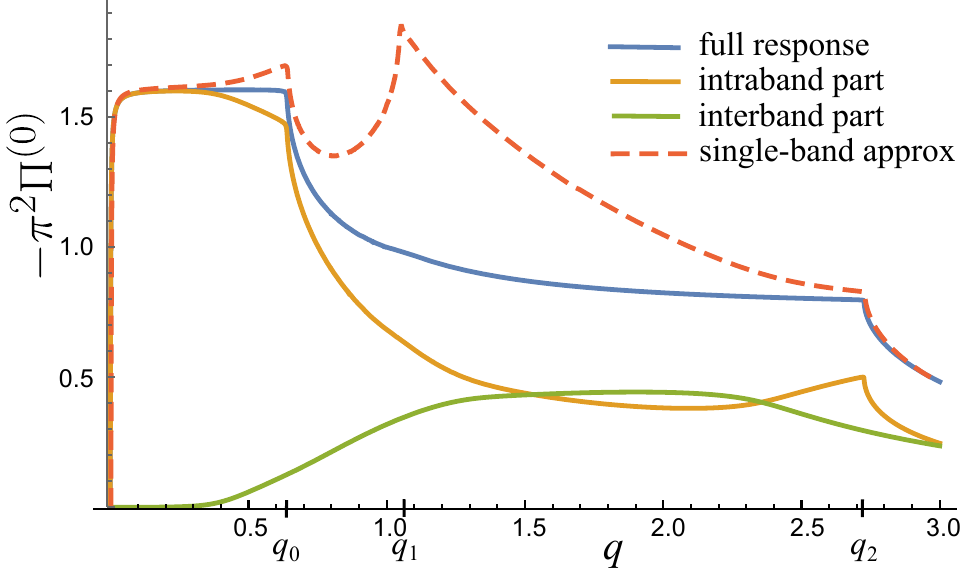}}
    \caption{The same as in Fig.~\ref{fig4}~(a) for $\varepsilon_F$ lying the upper part of the MHD. Calculations are performed for $\varepsilon_{F}=0.9$. The $q_1$-singularity, which is large in the case of the single-band approximation (dashed line), is actually absent (full lines).}
    \label{fig5}
\end{figure}

The evolution of the Lindhard polarization function with changing the Fermi energy $\varepsilon_F$ in the conduction band is shown in Fig.~\ref{fig3} and analyzed in more detail in Figs.~\ref{fig4},~\ref{fig5}. The amplitude of the $q_0$-singularity significantly exceeds the other two when $\varepsilon_F$ is in the lower part of the MHD, but it decreases rapidly with increasing $\varepsilon_F$ and finally disappears at $\varepsilon_F\approx\varepsilon_F^*$, where
\begin{equation}\label{eq.varepsilon*}
    \varepsilon_F^*=\frac{4}{5}\sqrt{1+\frac{9}{16}a^2-\frac{9}{64}a^4}\,.
\end{equation}
Interestingly, this critical value $\varepsilon_F^*$ corresponds to the condition $q_0(\varepsilon_F)=q_1(\varepsilon_F)$.

In contrast, the singularities at $q_1$ and $q_2$ increase with increasing $\varepsilon_F$ and become dominant in the upper part of the MHD.\@

The very large amplitude of the $q_0$-singularity at $q_0<<1$ is obviously due to the high density of states near the bottom of the MHD, but the quantum metric plays also an important role. To understand what other factors influence the features of the Lindhard function, we consider the contributions of intraband and interband transitions, as well as the effect produced by the quantum metric.

Calculations based on Eqs.~(\ref{eq.Pi(0)}), (\ref{eq.Pi_1}) and (\ref{eq.Pi_2}) show that the contributions of intraband and interband transitions to the total density response are, generally speaking, comparable in magnitude at $q\gtrsim 1$. The contribution of intraband transitions predominate only at $q\ll 1$. Figures~\ref{fig4}~(a) and~\ref{fig5} demonstrate this for certain values of $\varepsilon_F$.

The effect of the quantum metric can be easily seen by comparing the Lindhard function defined by Eq.~(\ref{eq.Pi(0)}), and especially its intraband part having singularities (see Eq.~(\ref{eq.Pi_1})), with a hypothetical Lindhard function $\Pi^{(0)}_{s-b}$ that would be obtained if we ignored the quantum metric, assumed $\mathcal{F}=1$, and discarded the other bands. One might say that this hypothetical Lindhard function corresponds to a ``single-band'' approximation. Results of this comparison are presented in Figs.~\ref{fig4}~(a, b) for the case when $\varepsilon<\varepsilon^*$ and the $q_0$-singularity is large.

In the case shown in Fig.~\ref{fig4}~(a), as can be seen, the quantum metric significantly suppresses the $q_0$-singularity. A more detailed study shows that the singularity suppression strongly depends on the Fermi energy. At a qualitative level, the amplitude of the singularity can be characterized by the height $h_{peak}$ of the peak of the function $\Pi^{(0)}(q)$ at $q=q_0$ relative to neighboring regions, where $\Pi^{(0)}$ changes smoothly. The peak height as a function of $\varepsilon_F$ is shown in Fig.~\ref{fig4}~(b) for the full Lindhard function $\Pi^{(0)}(q)$, its singular part $\Pi^{(0)}_{\rm{intra}}(q)$ that stems from the intraband transitions, and the ``single-band'' Lindhard function $\Pi^{(0)}_{s-b}(q)$. The effect of the quantum metric can be assessed qualitatively by comparing the “single-band” Lindhard function with the intra-band part. It is clear that the effect grows rapidly with $\varepsilon_F$. A comparison with the full Lindhard function, which contains a contribution from another band, would be less justified. But even in this case, the effect of the quantum metric is almost the same in magnitude and increases rapidly with $\varepsilon_F$. The quantum metric changes other singularities as well, but the most dramatic effect occurs when $\varepsilon_F$ is close to the bottom of the band. The $q_1$-singularity becomes very weak, and the $q_2$-singularity practically disappears.

In other telling case when $\varepsilon_F>\varepsilon_F^*$, the presence of a quantum metric leads to an even more interesting effect, as demonstrated in Fig.~\ref{fig5}. The Lindhard function $\Pi^{(0)}(q)$ does not have the $q_0$-singularity, although it exists for the ``single-band'' Lindhard function and even has a large amplitude. This happens because the metric-dependent function $\mathcal{F}(\bm{k}, \bm{q})$ in the integrand of Eq.~(\ref{eq.Pi_1}) eliminates the $q_0$-singularity of the integral.

The Lindhard function found here will be used further to calculate the electron density around an external charge. In the case of MHD, such a calculation should be carried out taking into account the e-e interaction.

\section{Electron density oscillations}\label{S_Friedel}

The electron density deviation, created by an external charge embedded at the point $r=0$ in the 2D layer of electrons is determined by the static density-density response function $\Pi(\bm{q})$ as follows
\begin{equation}
    \delta n(\bm{r})=\int \frac{d^2q}{(2\pi)^2} e^{i\bm{q r}} V_{ext}(q) \Pi(\bm{q})\,,
\end{equation}
where $V_{ext}(q)$ is the Fourier transform of an isotropic potential of the imbedded point charge. The response function $\Pi(\bm{q})$ is found within RPA~\cite{giuliani2008quantum} 
\begin{equation}\label{eq.RPA_Pi}
    \Pi(\bm{q})=\frac{2\Pi^{(0)}(\bm{q})}{1-2V_{ee}(q)\Pi^{(0)}(\bm{q})}\,,
\end{equation}
where $V_{ee}(q)$ is the Fourier transform of the e-e interaction potential. A factor of $2$ is added to account for both spin directions (up and down).

The oscillations of the electron density are found from the asymptotics of the density $\delta n(r)$ at $r\to\infty$. In the case of interest to us, when $\Pi(\bm{q})$ does not depend on the angular variable,
\begin{equation}\label{eq.n(r)_integral}
  \begin{split}
    \delta n(r)=\int \frac{dq q}{2\pi} V_{ext}(q) \Pi(q) J_0(qr)\bigg|_{r\to \infty} \\ \simeq \frac{e^{-i\pi/4}}{(2\pi)^{3/2}}\int_0^{\infty}\!dq q \Pi(q) V_{ext}(q) \frac{e^{iqr}}{\sqrt{qr}} +c.c.
  \end{split}
\end{equation}
The asymptotics of this integral is determined by the singularities of the function $\Pi(q)$ and properties of the integrand in the vicinity of the singular points. To calculate the asymptotics we use an approach similar to that proposed in Ref.~\cite{PhysRevB.81.205314} for 2D electrons with spin-orbit interaction. The idea is to replace the singular integrand with the sum of a smooth function, coinciding with the integrand in the singular points, and a piecewise smooth function that reproduces the singularities of the original integrand and vanishes at the singularity points.

From Eq.~(\ref{eq.RPA_Pi}) it is clear that the singularity points of $\Pi(q)$ coincide with those of $\Pi^{(0)}(q)$. They are $q_0$, $q_1$, and $q_2$. The function $\Pi(q)^{(0)}$ is presented in the form: $\Pi^{0}(q)=\widetilde{\Pi}^{(0)}(q)+\Delta\Pi^{(0)}(q)$, where $\widetilde{\Pi}^{(0)}(q)$ is a smooth function, coinciding with $\Pi^{(0)}(q)$ in the singularity points, and
\begin{multline}\label{eq.Pi_approx}
    \Delta\Pi^{(0)}(q)=\sum_{j=0}^2\left[B_{j, -}|q-q_j|^{\alpha_{j, -}}\theta(q_j-q) \right. \\
       +\left.B_{j, +}|q-q_j|^{\alpha_{j, +}}\theta(q-q_j)\right]\,,   
\end{multline}
with $j$ being the index of the singular points. Here $B_{j, \pm}$ and $\alpha_{j, \pm}$ are parameters that are defined in such a way that the behavior of the function $\Delta\Pi(q)$ closely approximates the function $\Pi^{(0)}(q)$ in the vicinity of the corresponding singularity points. This is achieved by numerically fitting the parameters. Thus, $B_{j, +}$ and $B_{j, -}$ characterize the amplitude of the singularity to the left and right of the point $q_j$, and $\alpha_{j, \pm}$ are the corresponding exponent. The smooth component $\widetilde{\Pi}^{(0)}(q)$ is of no importance in what follows, since the Fourier integral of it decreases much faster than the integrals generated by the singularities.

The asymptotics of the integral in Eq.~(\ref{eq.n(r)_integral}) is calculated by representing the function $\Pi(q)$ in the form: $\Pi(q)=\widetilde{\Pi}(q)+\Delta\Pi(q)$. This is easy to do using Eq.~(\ref{eq.RPA_Pi}) and taking into account that the singular term can be considered small. In this way we get
\begin{equation}
   \Delta\Pi(q)\approx \frac{2\Delta\Pi^{(0)}(q)}{[1-2V_{ee}(q)\Pi^{0}(q)]^2}\,. 
\end{equation}

As a result, we arrive at the following expression for the asymptotics of the density deviation:
\begin{multline}
    \delta n(r)\simeq \sum_{j=0}^2\left[\mathfrak{N}_{j, +}\frac{\cos\left(q_jr+\frac{\pi}{4}+\frac{\pi}{2}\alpha_{j, +}\right)}{r^{3/2+\alpha_{j, +}}}\right.\\-\left.\mathfrak{N}_{j, -}\frac{\cos\left(q_jr+\frac{\pi}{4}-\frac{\pi}{2}\alpha_{j, -}\right)}{r^{3/2+\alpha_{j, -}}}\right]\,,  
\end{multline}
where
\begin{equation}
    \mathfrak{N}_{j, \pm}=\frac{4}{(2\pi)^{3/2}}\frac{\sqrt{q_j}\,V_{ext}(q_j)}{[1-2V_{ee}(q_j)\Pi^{(0)}(q_j)]^2}\Gamma(\alpha_{j, \pm}+1) B_{j, \pm}\,,
\end{equation}
with $\Gamma(\alpha_{j, \pm}+1)$ being the $\Gamma$ function.

Thus, the FOs are formed by three modes with wave vectors $q_0$, $q_1$, and $q_2$. Moreover, each mode consists of two components with amplitudes $\mathfrak{N}_{j, \pm}$. They describe oscillations with the same wave vector and decay with distance according to a power law with different exponents. Mode interference leads to the appearance of different beat patterns depending on the amplitudes and phases of the components.

Now let us turn to a more specific case, when the external potential is created by a Coulomb center with a charge $Ze$ and the e-e interaction is also Coulomb. In dimensionless form
\begin{equation}
    V_{ext}(q)= \frac{Z Q}{q}\,,\quad V_{ee}(q)= \frac{Q}{q}\,,
\end{equation}
with $Q$ being the interaction parameter $Q=2\pi e^2/(\epsilon \sqrt{MB})$, and $\epsilon$ the effective background dielectric constant. In real materials, the value of $Q$ is scattered over a wide range due to changes in the dielectric constant and spectrum constants. In addition, $Q$ can also be considered as a model parameter of the interaction.

It this case the amplitudes of the FO modes are
\begin{equation}
    \mathfrak{N}_{j, \pm}=\frac{4 Z Q}{(2\pi)^{3/2}}\frac{q_j^{3/2}}{[q_j-2Q\Pi^{(0)}(q_j)]^2}\Gamma(\alpha_{j, \pm}+1) B_{j, \pm}\,.
\end{equation}

The asymptotic behavior of the modes with distance is determined by the parameters $\alpha_{j, \pm}$. The calculation shows that $\alpha_{0, -}=\alpha_{1, +}=\alpha_{2, +}=1$ and $\alpha_{0, +}\approx\alpha_{1, -}\approx\alpha_{2, -}\approx 1/2$. Therefore, the components with amplitudes $\mathfrak{N}_{0, -}$, $\mathfrak{N}_{1, +}$ and $\mathfrak{N}_{2, +}$ decay with distance as $r^{-2}$, and the components $\mathfrak{N}_{0, +}$, $\mathfrak{N}_{1, -}$ and $\mathfrak{N}_{2, -}$ decay as $r^{-5/2}$. Thus, there are six components in total, three of which dominate the asymptotics, while the other three can be significant near the charge $eZ$ where their amplitude can be large enough. As a result, a three-mode structure of the FOs is formed, the pattern of which is determined by the ratio of the amplitudes of the modes and their phase shifts. For this reason, the question of how the amplitudes $\mathfrak{N}_{j, \pm}$ depend on the Fermi energy, as it scans the MHD profile, and on the interaction parameter $Q$ is of crucial interest. 

\begin{figure}
    \centerline{\includegraphics[width=1.\linewidth]{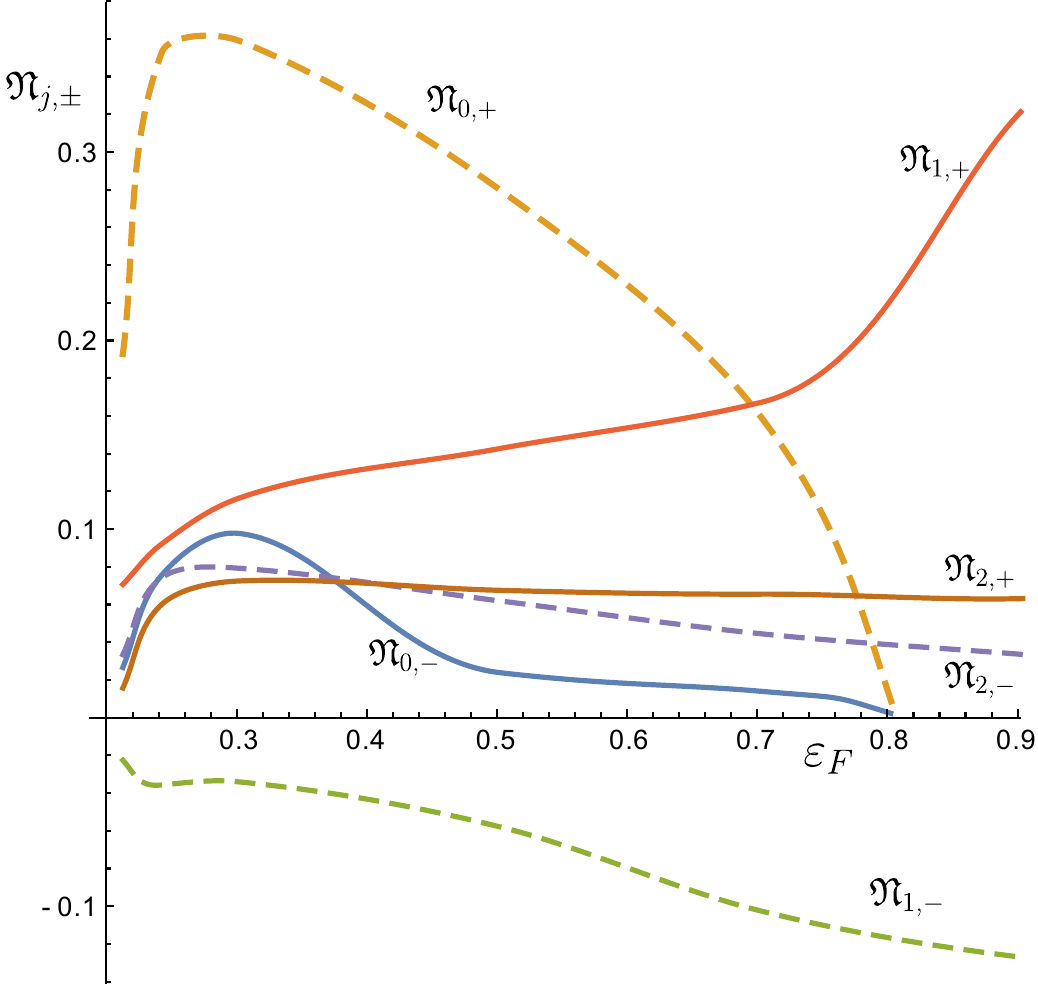}}
    \caption{Amplitudes $\mathfrak{N}_{j, \pm}$ of the FO components with three wave vectors $q_j$ as functions of the Fermi energy. The amplitudes $\mathfrak{N}_{0, -}, \mathfrak{N}_{1, +}, \mathfrak{N}_{2, +}$ refer to the components decaying with distance as $r^{-2}$, and $\mathfrak{N}_{0, +}, \mathfrak{N}_{1, -}, \mathfrak{N}_{2, -}$, shown by dashed lines, refer to the components decaying as $r^{-5/2}$. The parameters are $a=0.2$, $Q=1$.}
    \label{fig6}
\end{figure}

The results of the study of the dependence of the amplitudes on the Fermi energy are shown in Fig.~\ref{fig6} for a given value of the interaction parameter. The components $\mathfrak{N}_{0, +}$ and $\mathfrak{N}_{1, +}$ may appear to dominate, but this is only at a finite distance, since  $\mathfrak{N}_{0, +}$ component decreases faster as $r$ increases and so $\mathfrak{N}_{1, +}$, $\mathfrak{N}_{0, -}$ and $\mathfrak{N}_{2, +}$ components dominate asymptotically in various combinations depending on $\varepsilon_F$. The amplitudes of these main components exhibit completely different behavior depending on the Fermi energy. The general pattern is that the $q_0$-mode is significant in the low-energy part of the MHD, while the $q_1$- and $q_2$-modes dominate in the upper part. In addition, the amplitude of the $q_0$-mode reaches a maximum in the lower part of the MHD.\@ It is important to note that the height of the maximum increases rapidly with decreasing the interaction parameter.

\begin{figure}
    \centerline{\includegraphics[width=1.0\linewidth]{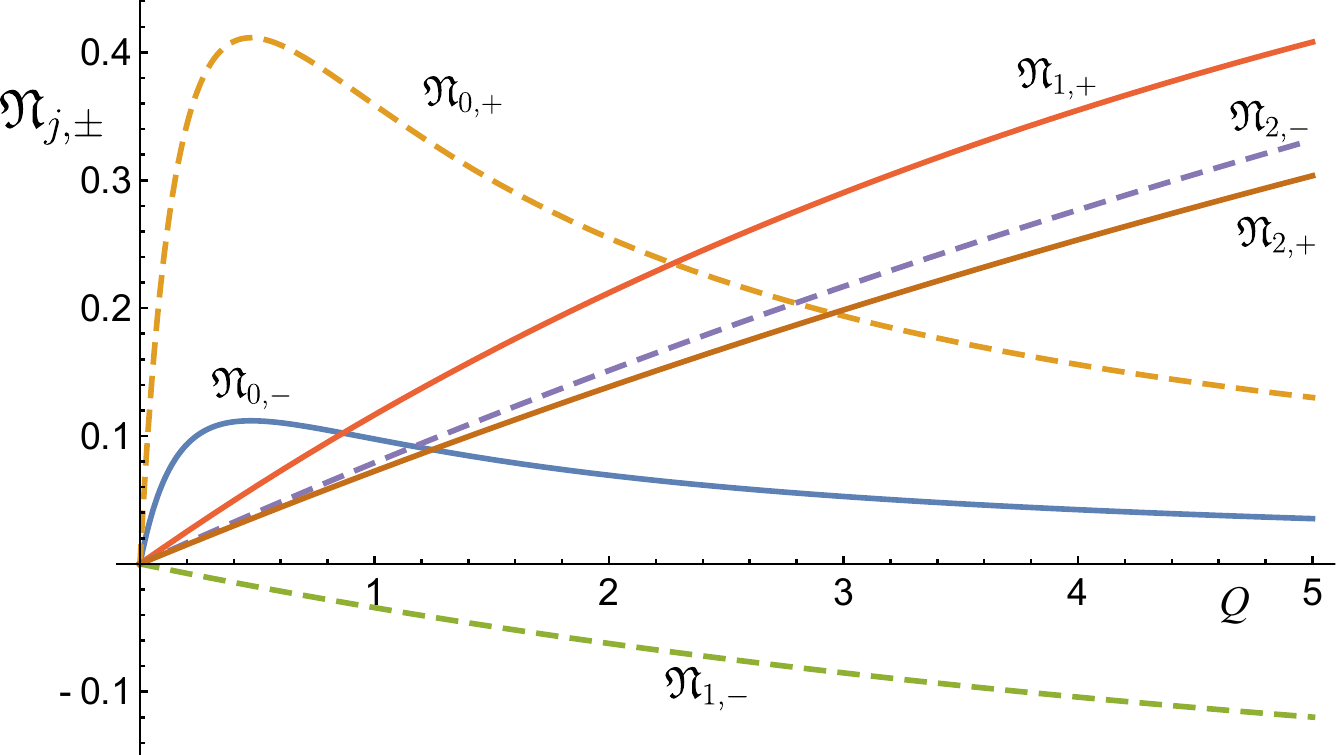}}
    \caption{Amplitudes of all six components of the FOs as functions of $Q$. The line designations are the same as in Fig.~\ref{fig6}. The parameters are $a=0.2$, $\varepsilon_F=0.3$.}
    \label{fig7}
\end{figure}

Another interesting feature of the FOs, characteristic of the MHD, is the dependence of the amplitudes of all three modes on the interaction parameter $Q$. This dependence is shown in Fig.~\ref{fig7} for a given $\varepsilon_F$. There is also a fundamental difference in the behavior of the inter-contour $q_0$-modes and the intra-contour $q_1$- and $q_2$-modes. The amplitudes of the $q_0$-modes reach a maximum at $Q=Q_0(\varepsilon_F)$, where $Q_0$ is estimated as
\begin{equation}
    Q_0(\varepsilon_F)=\frac{q_0(\varepsilon_F)}{2 \Pi^{(0)}(q_0(\varepsilon_F))}\,.
\end{equation}

At $Q>Q_0$, the amplitudes $\mathfrak{N}_{0, \pm}$ decrease with increasing $Q$. On the contrary, the amplitudes of the $q_1$- and $q_2$-modes increase monotonically. And it is important to note that for small $Q$ the amplitudes of the $q_0$-mode always exceed all the others as $\varepsilon_F$ approaches the MHD bottom. 

The tendency to zero for all amplitudes as $Q$ goes to zero is not very interesting, since it is trivially explained by the tendency to zero of the disturbing potential.

The interference of the FO modes gives rise to a variety of beat patterns of the electron density. As an illustration, we present only two limiting cases where $\varepsilon_F$ lies near the bottom and near the top of the MHD.\@ At the bottom, Fig.~\ref{fig7}a, the beat pattern is determined mainly by the mixing of the $q_0$- and $q_1$-modes. Near the top, Fig.~\ref{fig7}b, the beats are formed by the $q_1$- and $q_2$-modes.

\begin{figure}
    \centerline{\includegraphics[width=1.\linewidth]{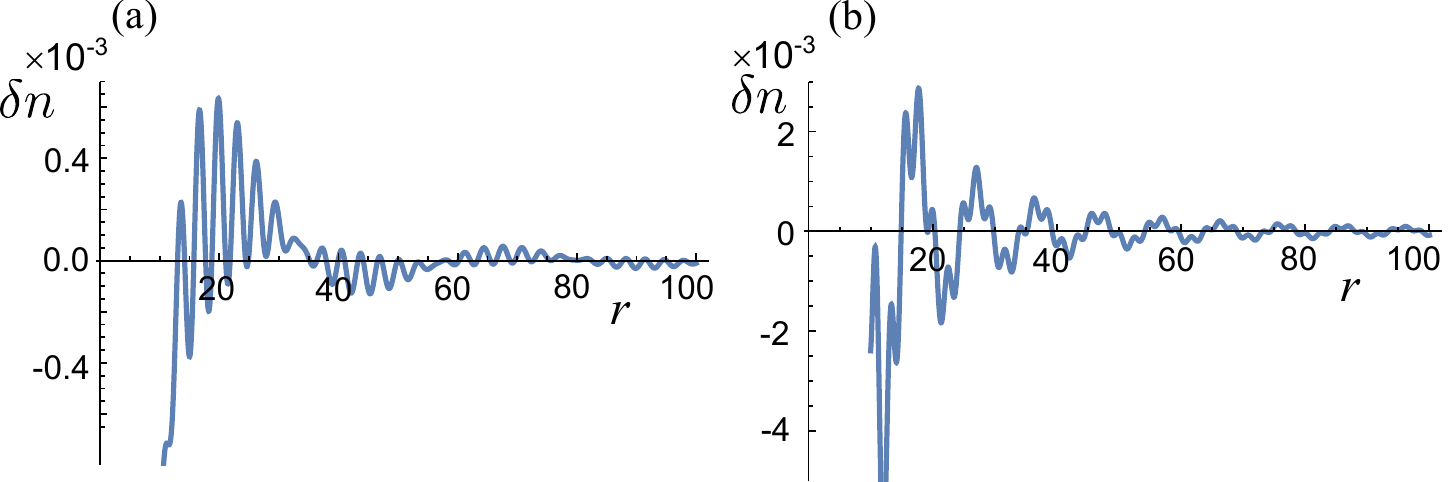}}
    \caption{Beating of the FOs when the Fermi energy is close to (a) the bottom and (b) the top of the MHD.\@ For panel (a) the parameters are:  $a=0.2$, $\varepsilon_F-\varepsilon_c=0.04$ and $Q=1$; for panel (b), $a=0.2$, $\varepsilon_t-\varepsilon_F=0.1$ and $Q=5$.}
    \label{fig8}
\end{figure}

\section{Discussion and concluding remarks}\label{S_discuss}

Interest in FOs in 2D materials with MHD is due to a bunch of non-trivial features inherent in such dispersion: Van Hove singularity, the presence of two Fermi contours in the energy region between the bottom and the top of MHD, hybridization of Bloch orbitals, due to which non-trivial quantum-geometric properties are formed, the presence of states with a negative effective mass.

As a specific model of MHD, we used the BHZ model according to which MHD is formed as a result of $s p^3$ hybridization, when the electron and hole subbands are inverted.

The presence of two Fermi contours, the space between which is filled with electrons, gives rise to three singularities of the Lindhard polarization function. In addition to the two Kohn anomalies at $q_1=2k_{F1}$ and $q_2=2k_{F2}$, arising from electron transitions within each of the two Fermi contours, there is another singularity at the wave vector $q_0=k_{F2}-k_{F1}$, corresponding to transitions between the Fermi contours. This last singularity turns out to be much larger than the other two in amplitude when the Fermi energy is near the bottom of the MHD.\@ This is obviously due to the divergence of the density of states at the bottom of the band.

Already at this stage, an effect caused by the quantum geometry of the band states is revealed. The presence of the overlap function representing the quantum metric leads to disappearance of the $q_0$-singularity when $\varepsilon_F>\varepsilon_F^*$ (in terms of Fermi wave vectors, this condition corresponds to $q_0>q_1$). Quantum geometry also affects the two other singularities at $q_1$ and $q_2$. They are significantly reduced in magnitude, especially the $q_2$-singularity, which practically disappears when $\varepsilon_F$ is close to the MHD bottom.

Electron density oscillations around a point defect with a Coulomb potential are calculated taking into account the e-e interaction within the RPA.\@ We found the amplitudes of all three modes of the FOs with wave vectors $q_0$, $q_1$ and $q_2$. Each mode has two components decaying with the distance as $\sim r^{-2}$ and $\sim r^{-5/2}$. Interference of the modes results in a variety of patterns of spatial beats of the electron density, which are determined by the ratio of the amplitudes of the FO components and, to a lesser extent, their phases. The amplitudes of the modes change in a manner specific to each mode depending on the Fermi energy. For the $q_0$-mode, the amplitudes $\mathfrak{N}_{0, \pm}$ of both components reach their maximum near the band bottom and then decrease with increasing $\varepsilon_F$. The amplitudes $\mathfrak{N}_{1, \pm}$ and $\mathfrak{N}_{2, \pm}$ of other two modes, on the contrary, increase monotonically with increasing $\varepsilon_F$ starting from small values, and the amplitudes of the $q_1$-mode exceed the amplitudes of the $q_2$-mode. Thus, the $q_0$-mode plays an important role when $\varepsilon_F$ is in the lower part of the MHD.\@ In this case, the FOs are formed mainly by the $q_0$- and $q_1$-modes. In the upper part of the MHD, the $q_1$- and $q_2$-modes become decisive.

Another interesting feature arises in the behavior of the $q_1$-mode as a function of $\varepsilon_F$ in the upper part of the MHD.\@ In this case, the effective mass is negative, since the $q_1$-mode is created by electron transitions within the low-$k$ branch of the MHD.\@ The amplitude $\mathfrak{N}_{1, +}$ starts to grow rapidly as $\varepsilon_F$ approaches the top of the MHD.\@ The effect strongly increases with the interaction parameter $Q$. This behavior of the $q_1$-mode clearly shows that some electrons are effectively attracted to the repulsive center.

The interaction between electrons plays an important role in shaping the spatial structure of the FOs, since the e-e interaction changes the amplitudes of the modes in different way. The amplitudes of the $q_0$-mode, as a function of the interaction parameter $Q$, has a maximum at a certain value of $Q = Q_0$ and then decreases with increasing $Q$. The amplitudes of the $q_1$- and $q_2$-modes, on the contrary, monotonically increase with increasing $Q$, starting from small values. As a result, the $q_0$-mode is significant and even predominates when the interaction is relatively weak, but an increase in $Q$ leads to the $q_1$- and $q_2$-modes becoming predominant.

The main conclusion of this paper, that the structure of FOs is determined not only by the dispersion of band electrons but also by the quantum metric of the band states, is quite general and applicable to many systems. Such a universal role of the quantum metric is due to the fact that it determines the distance between neighboring Bloch states and thus significantly affects the susceptibility, which is defined as the integral over the Bloch states. But for the effect of the quantum metric to be significant, it must change sufficiently strongly depending on the wave vector. It should be noted that the Berry curvature is not of fundamental importance and may even be absent.

An example of such a system, in which the Berry curvature is absent, is the well-known model system of a 2D electron gas with spin-orbit interaction. It also has a Mexican hat dispersion and, accordingly, two Fermi contours $k_{F1}$ and $k_{F2}$, but, as calculations show~\cite{PhysRevB.74.045307}, there is only one Kohn anomaly at $q=k_{F2}-k_{F1}$, corresponding to transitions between the contours. Intraband anomalies at $q=2k_{F1}$ and $q=2k_{F2}$ are absent, as we found out, precisely because of the metric.

The Berry curvature in such a system appears in the presence of a perpendicular magnetic field. As is easy to show, in this case the metric changes significantly, which leads to corresponding changes in the structure of FOs~\footnote{These results will be published in more detail elsewhere.}. Thus, this is a suitable system for implementing the quantum metric effect which can be realized in a large number of materials with strong spin-orbit interaction~\cite{manchon2015new,Premasiri_2019}. The quantum metric effect can manifest itself in a range of energies that is determined by the magnitude of the spin-orbit splitting.

The system studied in this paper has richer possibilities, since it explicitly contains a mechanism of strong spin-orbit interaction due to $sp^3$ hybridization and nontrivial topological properties due to band inversion. Systems of this type with MHD have been implemented in materials such as Bi$_{1.1}$Sb$_{0.9}$Te$_2$S~\cite{PhysRevB.101.121115}, but the greatest attention has been attracted by inverted quantum wells InAs/GaSb with controlled band inversion, which have a sufficiently large band gap~\cite{PhysRevLett.100.236601,PhysRevResearch.4.L042042}. The characteristic energy at which the pattern of FO beats is restructured due to the interplay of the quantum metric and inter-contour transitions is estimated at the level of $\varepsilon^*$, Eq.~(\ref{eq.varepsilon*}) for a given hybridization parameter.

The overall conclusion is that the atomic orbital structure and quantum metric of the electronic states manifest themselves in specific features of the FOs, and the analysis of the FO modes provides valuable information about the electronic states. We believe that the FO features found here can be directly observed by STM imaging~\cite{doi:10.1126/science.1142882,Chen_2017} of the distribution of local density of states around a repulsive point defect or other surface imperfections.

\begin{acknowledgments}
This work was carried out in the framework of the state task for the Kotelnikov Institute of Radio Engineering and Electronics.
\end{acknowledgments}
        
\bibliography{friedel_oscil_mex-hat}

\end{document}